\documentclass[a4paper,11pt]{article}
\usepackage{graphicx} % Required for inserting images
\usepackage{epsfig,latexsym,amsfonts,amsmath,amsthm,amssymb,amsbsy,multirow,slashed,wasysym,textcomp,subfigure,hyperref,wrapfig,datetime,comment,mathtools,cancel,dsfont,diagbox}

\newcommand{\badat}{\begin{alignedat}}
\newcommand{\eadat}{\end{alignedat}}

\usepackage{color}

\newcommand{\Dcal}{\mathcal{D}}

\newcommand{\scri}{\mathcal{I}}

\newcommand{\varep}{{\overline{\epsilon}(\hat{x})}}
\newcommand{\vare}{{\overline{\epsilon}}}

\usepackage[sorting=none,style=numeric-comp,maxbibnames=99]{biblatex}

\numberwithin{equation}{section}

\begin{document}

\begin{titlepage}
  \thispagestyle{empty}
 
  \bigskip\ \bigskip\ \bigskip
  \bigskip\ \bigskip\ \bigskip
  \begin{center}

        \baselineskip=12pt {\LARGE \scshape{
        
        Exact Infrared Triangle in Massless sQED with Long-range Interactions

        }}
  
      \vskip1.5cm 

   \centerline{
        Sangmin Choi\footnotemark[1],
        Ameya Kadhe\footnotemark[2],
        and Andrea Puhm\footnotemark[3]
   }

\bigskip\bigskip

\centerline{\em {Institute for Theoretical Physics, University of Amsterdam,}}\smallskip
\centerline{\em {PO Box 94485, 1090 GL Amsterdam, The Netherlands}}

\smallskip

\bigskip\bigskip
 
\end{center}

 \begin{abstract}
   The logarithmic soft photon theorem in four spacetime dimensions encodes an infinite-dimensional asymptotic symmetry which acts on massive matter as a divergent superphaserotation. Here we extend this result to massless matter which is both more subtle and surprising. We derive the charges associated to divergent superphaserotations and show that the logarithmically divergent charge exactly vanishes to all orders in the electromagnetic coupling. This is in agreement with the vanishing of the classical logarithmic soft photon theorem which is one-loop exact.  Special care is required for massless matter due to potential collinear divergences which, as we show, do however not affect the superphaserotation charge. We furthermore compute the infrared corrections to the charge associated to the subleading tree-level soft photon theorem.
  As a corollary of our result, we find that the tail to the velocity kick memory due to the long-range interactions between soft electromagnetic radiation and massless matter vanishes. 
 \end{abstract}

\footnotetext[1]{Email: s.choi@uva.nl}
\footnotetext[2]{Email: a.m.kadhe@uva.nl}
\footnotetext[3]{Email: a.puhm@uva.nl}

\end{titlepage}

%\maketitle
\tableofcontents

\newpage

%%%%%%%%%%%%%%%%%%%%%%%%%%%%%%%%%%
\section{Introduction}
%%%%%%%%%%%%%%%%%%%%%%%%%%%%%%%%%%

Over the past decade, a trove of insights into the infrared structure of gauge theories has been obtained which explains the universality of low-energy and large-distance properties of scattering by underlying symmetries. 
Scattering amplitudes exhibit factorization when low-energy, or {\it soft}, gauge bosons are emitted, radiation leads to memory effects at large distances, and both effects can be traced to the asymptotic symmetries of gauge theories in (asymptotically) flat spacetimes. 

The simplest example of such an {\it infrared triangle} consists of the soft photon theorem in QED, the electromagnetic kick memory effect and large gauge symmetry which acts on matter as a local enhancement of global phase rotations, or, {\it superphaserotations}; it manifests itself in the conservation of a Noether charge at every angle whose global version is the familiar U(1) charge of QED \cite{He:2014cra, Lysov:2014csa}. 
Analogous infrared triangles have been found in other theories including gravity \cite{Strominger:2013jfa}, non-Abelian gauge theory \cite{He:2015zea} and supersymmetric theories \cite{Dumitrescu:2015fej}. The study of the infrared structure has also been extended to subleading orders in the low-energy expansion \cite{Lysov:2014csa, Campiglia:2015kxa, Kapec:2014opa,Campiglia:2016efb} which translates to weaker large-distance fall-offs of the radiative field in the corresponding memory effect analyses and to overleading orders in a large-distance expansion of the associated symmetry parameters \cite{Campiglia:2016hvg,Campiglia:2016efb,Choi:2024ygx,Choi:2024ajz}.

The long-range nature of electromagnetic and gravitational interactions turns out to modify all these infrared triangles beyond the leading one: early/late-time divergences were shown to dress the matter fields \cite{Campiglia:2019wxe} by logarithmically divergent phases and this behavior dominates over the free field asymptotics!   Only recently have these effects been accounted for in the discussion of the infrared triangles \cite{Choi:2024ygx, Choi:2024mac, Choi:2024ajz} and so far only in (scalar) QED and gravity with massive matter. Our goal in this work is to connect the logarithmic soft photon theorem in massless scalar QED to an asymptotic symmetry analysis. A corresponding analysis for gravity coupled to massless matter is in progress \cite{Choi-Kadhe-Puhm}.

\paragraph{Power-law soft photon theorems.} 
The leading soft photon theorem states that the scattering amplitude for $N+1$ particles, one of which is a photon with momentum $k^{\mu}=\omega q^{\mu}$, factorizes in the $\omega\to 0$ limit into the scattering amplitude of the $N$ hard particles with momenta $p_i$,
\begin{equation}\label{Weinberg Soft Theorem}
\lim_{\omega\to 0}\omega\mathcal{M}_{N+1}(\{p_{i}\};(\omega,q,\ell))=S_{-1}(\{p_{i}\};(q,\ell))\mathcal{M}_{N}(\{p_{i}\}),
\end{equation}
where the soft factor $S_{-1}$ is given by
\begin{equation}
S_{-1}=e\sum_{i=1}^{N}\frac{Q_{i}p^{\mu}_{i}\varepsilon_{\mu}}{q\cdot p_{i}}. \label{Weinberg soft factor}
\end{equation}
The leading soft theorem is universal as it depends only on the charges and momenta of the external hard particles involved in the scattering process. 

Assuming the validity of a power-law soft expansion, there exists a subleading soft photon theorem\footnote{The operator $\lim_{\omega\to 0}(1+\omega\partial_{\omega})$ acts as a projector that extracts the $\omega^0$ term from the soft expansion.} 
\begin{equation}\label{Low Soft Theorem}
\lim_{\omega\to 0}(1+\omega\partial_{\omega})\mathcal{M}_{N+1}(\{p_{i}\};(\omega,q,\ell))=S_{0}(\{p_{i}\};(q,\ell))\mathcal{M}_{N}(\{p_{i}\}),
\end{equation}
with the soft factor given by
\begin{equation}
S_{0}=-ie\sum_{i=1}^{N}\frac{Q_{i}q_{\mu}\varepsilon_{\nu}J^{\mu\nu}_{i}}{q\cdot p_{i}}.\label{Low soft factor}
\end{equation}
Unlike the leading soft factor \eqref{Weinberg soft factor}, the subleading soft factor $S_{0}$ is not universal. It receives corrections from a (finite) set of effective field theory operators \cite{Elvang:2016qvq} with three-point interactions as well as loop corrections \cite{Bern:2014oka,Laddha:2018myi,Laddha:2018vbn,Sahoo:2018lxl}. These loop corrections enter at $O(\ln\omega)$ and so are overleading compared to the subleading tree-level soft factor $\sim \omega^0$ which the presence of the logarithm renders ambiguous.

\paragraph{Logarithmic soft photon theorems.} In (massive) scalar QED, it has been conjectured \cite{Karan:2025ndk} that given a well-defined soft expansion to all orders in the loop expansion, the ratio between the $(N+1)$-particle amplitude including a soft photon and the corresponding $N$-particle amplitude without that soft photon is infrared-finite. This allows us to define the loop-corrected expansion of the soft factor as
\begin{align}
\frac{\mathcal{M}_{N+1}(\{p_{1},\cdots, p_{N}\};(\omega,q,\ell))}{\mathcal{M}_{N}(p_{1},\cdots, p_{N})}=\sum_{n=-1}^{\infty}\omega^{n}(\ln\omega)^{n+1}S_{n}^{(\ln\omega)}+\dots .
\label{Conjectured structure of the logarithmic soft theorems}
\end{align}
The logarithmic soft theorem in theories with massive matter receives contributions from classical infrared and quantum loop effects \cite{Sahoo:2018lxl}
\begin{equation}
S_{0}^{(\ln\omega)}=S_{0,\text{classical}}^{(\ln\omega)}+\Delta S_{0,\text{quantum}}^{(\ln\omega)},
\end{equation}
given by
\begin{equation}
\label{Sclassical}
S_{0,\text{classical}}^{(\ln\omega)}=-ie^{3}\sum_{i=1}^{N}\frac{\varepsilon_{\mu}q_{\nu}Q_{i}}{q\cdot p_{i}}\sum_{j\neq i\atop\eta_{i}\eta_{j}=1}\frac{Q_{i}Q_{j}}{4\pi}\frac{p_{i}^{2}p_{j}^{2}\Big[p_{i}^{\mu}p_{j}^{\nu}-p_{j}^{\mu}p_{i}^{\nu}\Big]}{\Big[\big(p_{i}\cdot p_{j}\big)^2-p_{i}^{2}p_{j}^{2}\Big]^{3/2}},
\end{equation}
and
\begin{equation}
\label{Sln0sQEDquantum}
\badat{2}
   S_{0,\text{quantum}}^{(\ln\omega)}&=  e^3\sum_{i=1}^N\frac{\varepsilon_\mu q_\nu Q_i}{ q\cdot p_i}\left(p^\mu_i\partial^\nu_{p_i}-p^\nu_i\partial^\mu_{p_i}\right)\\
    &\quad \sum_{j\neq i} \frac{Q_iQ_j}{8\pi^2} \frac{(p_i\cdot p_j)}{\sqrt{(p_i\cdot p_j)^2-p_i^2 p_j^2}}\ln\left(\frac{p_i\cdot p_j+\sqrt{(p_i\cdot p_j)^2-p_i^2p_j^2}}{p_i\cdot p_j-\sqrt{(p_i\cdot p_j)^2-p_i^2p_j^2}}\right).
\eadat
\end{equation}

An asymptotic symmetry explanation of the classical soft theorem was given in \cite{Choi:2024mac} which derived a `soft' charge that maps on to the soft photon insertion in the S-matrix and a `hard' charge whose action on the matter fields produces the classical soft factor. The conservation of the sum of these logarithmic soft and hard charges between past and future boundaries is exact in the electromagnetic coupling and yields the classical logarithmic soft photon theorem which is one-loop exact.

Here we would like to explain the symmetry underlying the classical logarithmic soft photon theorem in theories with massless matter. However, it is easy to see that the (naive\footnote{The absence of potential collinear divergences is shown in appendix \ref{Appendix on collinear divergences}.}) massless limit of \eqref{Sclassical} vanishes. This points towards the classical one-loop corrected scattering structure of massless QED being trivial: long-range effects do not correct the subleading tree-level structure. Moreover the subleading tree-level soft theorem is not rendered ambiguous as they would have been in the presence of logarithms. The only non-vanishing contribution comes from quantum effects, but how to explain them from an asymptotic symmetry analysis is as of yet unclear, and we will not attempt to do so here\footnote{Attempts in this direction were made in \cite{Campiglia:2019wxe}.}. In an asymptotic symmetry derivation of the classical logarithmic soft theorem with massless matter we thus have to show that both soft and hard charges exactly vanish to all orders in the electromagnetic coupling!

In this work, we construct an asymptotic phase space that accounts perturbatively for the long-range interactions between the electromagnetic field and the massless matter. This is significantly more tricky than for massive matter since radiation and massless matter reach the same spacetime boundary and the radiative and Coulombic corrections from interactions mix.

As ansatz for the relevant symmetry transformation we consider the same linearly divergent superphaserotation that is responsible for the subleading tree-level soft photon theorem. This is motivated by the corresponding derivation of the logarithmic soft photon theorem in massive scalar QED. Indeed we will see here that this ansatz reproduces the leading and subleading tree-level soft theorems in massless scalar QED in addition to explaining the exact vanishing of logarithmic soft photon theorem from symmetry\footnote{
See also the related work \cite{AtulBhatkar:2019vcb}.}. 
Moreover, we determine the long-range correction to the charge associated to the subleading soft theorem \cite{Campiglia:2016hvg}, at higher order in the electromagnetic coupling which corresponds to the one-loop correction to Low's subleading soft photon theorem. Care has to be taken with potential collinear divergences, which we show do, however, not affect our classical analysis.

Finally, a corollary of our result is that the infrared triangle in massless scalar QED with long-range interactions has all the three corners evaluated to \textit{zero}. That is, there is no non-trivial classical (leading) logarithmic soft photon theorem, there is no classical tail to the velocity-memory, and there are no non-trivial logarithmic charges. This result holds to all orders in the electromagnetic coupling.\\

This work is organized as follows. In section \ref{AsymptoticData} we discuss the asymptotic behavior of massless scalar and gauge fields and the infrared corrections due to interactions. We state in section \ref{SuperphaserotationSymmetry} large gauge transformation given by a linearly divergent superphaserotation which is then used in section \ref{Superphaserotation charge section} to compute the asymptotic charge associated to the classical logarithmic soft photon theorem. In section \ref{TailMemory} we deduce the tail to the electromagnetic kick memory for massless matter. We conclude in section \ref{Conclusion} with a discussion of our result.

%%%%%%%%%%%%%%%%%%%%%%%%%%%%%%%%%%
\section{Asymptotic data at null infinity}
\label{AsymptoticData}
%%%%%%%%%%%%%%%%%%%%%%%%%%%%%%%%%%
In the following we will discuss the asymptotic fall-offs for the charged scalar matter and the gauge field at the future boundary, but a similar analysis can be repeated at the past boundary.
We consider a massless complex scalar $\phi$ coupled minimally to a gauge field $A_\mu$ with
field strength $ F_{\mu\nu}=\partial_\mu A_\nu-\partial_\nu  A_\mu$ described by the action
\begin{equation}
    S
    =
        -\int d^4x\left\{
            \left[(\nabla_\mu - ieA_\mu)\phi\right]^*
            \left[\nabla^\mu-ieA^\mu\right]\phi
            + \frac14 F_{\mu\nu}F^{\mu\nu}
        \right\}.
\end{equation}
The equations of motion are
\begin{equation} \label{Field Equations for phi and A}
      (\nabla_\mu - ie A_\mu)(\nabla^\mu - ie A^\mu) \phi=0,
   \qquad \nabla^\nu F_{\mu\nu}=j_\mu,
\end{equation}
with matter current given by
\begin{equation}
 j_\mu=i e\phi (\nabla_\mu + ie A_\mu) \phi^* +{\rm c.c.}\,.
\end{equation}
We will work in Lorenz gauge $\nabla^\mu A_\mu=0$ for which Maxwell's equation reduces to $\nabla^2 A_\mu=-j_\mu$.
We use retarded Bondi coordinates, in terms of which the flat metric is
\begin{align}
\text{d}s^{2}&=-\text{d}u^{2}-2\text{d}u\text{d}r+r^{2}\gamma_{AB}\text{d}x^{A}\text{d}x^{B}.
\end{align}
Here $u=t-r$ is the retarded time and $\gamma_{AB}$ is the metric on the unit sphere $S^2$. 
In these coordinates the Lorenz gauge condition takes the form
\begin{equation}
\label{Lorenzscri}
    \partial_r\left(r^2A_u\right) -\partial_r\left(r^2A_r\right)+r^2\partial_u A_r- D^C A_C=0,
\end{equation}
where $D_C$ is the covariant derivative on $S^2$, and Maxwell's equations are given by
\begin{equation}
\label{Maxwellscri}
\badat{3}
j_u&=\Big[-\partial_r^2+2\partial_r \partial_u+\frac{2}{r} (\partial_u-\partial_r) -\frac{1}{r^2}D^2\Big]  A_u,\\
j_r&=\Big[-\partial_r^2+2\partial_r \partial_u+\frac{4}{r}(\partial_u-\partial_r) -\frac{1}{r^2}(D^2+2)\Big] A_r+\Big[\frac{2}{r}\partial_r+\frac{2}{r^2}\Big] A_u
,\\
j_A&=\Big[-\partial_r^2+2\partial_r \partial_u -\frac{1}{r^2}(D^2-1)\Big]  A_A+\frac{2}{r}D_A(A_{u}- A_{r}).
\eadat
\end{equation}

%%%%%%%%%%%%%%%%%%%%%%%%%%%%%%%%%%
\subsection{Free fields}
%%%%%%%%%%%%%%%%%%%%%%%%%%%%%%%%%%

We begin with a brief review of the asymptotic behavior of free fields.

%%%%%%%%%%%%%%%%%%%%%%%%%%%%%%%%%%
\subsubsection*{Radiation}
%%%%%%%%%%%%%%%%%%%%%%%%%%%%%%%%%%

The {\it free} Maxwell field in Minkowski space can be expressed as 
\begin{equation}\label{Acalfree}
    A^{\rm free}_\mu(x)=\int \frac{d^3k}{(2\pi)^32\omega_k}\left[a_{\mu}(\vec k)e^{ik\cdot x}+{a}_{\mu}(\vec k)^\dagger e^{-ik\cdot x}\right],
\end{equation}
with $\omega_k=k^0=|\vec k|$, $k\cdot x= -\omega_k t+\vec k \cdot \vec x$ and $a_{\mu}(\vec k)=\sum_{\lambda=\pm} \varepsilon^{\lambda*}_\mu a_\lambda(\vec k)$. Its asymptotic behavior is obtained from a saddle point analysis with saddle point $\vec k=\omega_k\hat{x}$, where we use the notation $\hat{x}$ to denote the unit three-dimensional vector. The $r\to \infty$ limit (at fixed $u$) is
\begin{equation}\label{Acalfreesaddle}
    A^{\rm free}_\mu(x)=-\frac{i}{8\pi^2r}\int_0^\infty d\omega_k \left[a_\mu(\omega_k,x^A)e^{-i\omega_k u}-a_\mu(\omega_k,x^A)^\dagger e^{i\omega_k u}\right]+O(1/r^2),
\end{equation}
which, after mapping the Cartesian components to Bondi coordinates, yields
\begin{equation}
    A^{\rm free}_u=O(r^{-1}),\quad A^{\rm free}_r=0,\quad A^{\rm free}_C=O(r^0).
\end{equation}
The free data of the photon field at $\scri^+$ is given by 
\begin{equation}
\label{calAClarger}
A_{C}^{\text{free}}(u,r,\hat{x})\stackrel{r\to \infty}{=} A^{\rm free}_C(u,\hat{x})+\dots\,,
\end{equation}
whose early/late times $u\to \pm\infty$ limit takes the form
\begin{align}
\label{ACufalloff_free}
        A_C^{\rm free}(u,\hat{x})
        &\overset{u\to\pm\infty}=
            A_C^{(0),\pm}(\hat{x})
            + O\Big(\frac{1}{ u^{\#}}\Big)
\end{align}
where  $\#>1$.

%%%%%%%%%%%%%%%%%%%%%%%%%%%%%%%%%%
\subsubsection*{Massless matter}
%%%%%%%%%%%%%%%%%%%%%%%%%%%%%%%%%%

The \textit{free} complex scalar field in Minkowski space can be expressed as
\begin{equation}
\phi^{\text{free}}(x)=\int\frac{d^{3}k}{(2\pi)^{3}2\omega_{k}}\left[a(\vec{k})e^{ik\cdot x}+b(\vec{k})^{\dagger}e^{-ik\cdot x} \right],
\end{equation}
with $\omega_{k}=k^{0}=|\vec{k}|,$ $k\cdot x=-\omega_{k}t+\vec{k}\cdot \vec{x}$. Performing again a saddle point analysis with saddle point $\vec{k}=\omega_{k}\hat{x}$ gives its asymptotic behavior for the $r\to\infty$ limit (at fixed $u$)
\begin{align}
\phi^{\text{free}}(x)=-\frac{i}{8\pi^{2}r}\int_{0}^{\infty}d\omega_{k}\left[a(\omega_{k},\hat{x})e^{-i\omega_{k}u}-b^{\dagger}(\omega_{k},\hat{x})e^{i\omega_{k}u}\right]+O(1/r^{2}).
\end{align}
Thus the large $r$ behaviour of the free scalar field is $\phi^{\text{free}}(x)=O(r^{-1})$ and we write
\begin{align}
\phi^{\text{free}}(u,r,\hat{x})\overset{r\to\infty}{=}\frac{1}{r}\phi_{\text{free}}^{1}(u,\hat{x})\hspace{1mm}+o(1/r).
\end{align}
The free field $U\left(1\right)$ current is given by
\begin{align}
j_{\mu}^{\text{free}}=ie(\phi_{\text{free}}\partial_{\mu}\phi_{\text{free}}^{*})+\text{c.c.},
\end{align}
whose components in Bondi coordinates have the following large $r$ falloff
\begin{equation}
\label{j-free, r-falloff}
j_{u}^\text{free}=O\left(r^{-2}\right)
,\qquad
j_{r}^\text{free}=O\left(r^{-4}\right)
,\qquad
j_{C}^\text{free}=O\left(r^{-2}\right). 
\end{equation}

%%%%%%%%%%%%%%%%%%%%%%%%%%%%%%%%%%
\subsection{Interacting fields}
%%%%%%%%%%%%%%%%%%%%%%%%%%%%%%%%%%

From the free-field fall-offs we can now determine via the equations of motion the corrections due to the interactions between radiation and matter.

%%%%%%%%%%%%%%%%%%%%%%%%%%%%%%%%%%
\subsubsection*{Radiative and Coulombic gauge modes}
%%%%%%%%%%%%%%%%%%%%%%%%%%%%%%%%%%

The free current, via the Maxwell equations \eqref{Maxwellscri}, sources a Coulombic potential which will have both powers and logarithms in $r$ in the large-$r$ expansion. In what follows, we use the notation $\overset{k,n}{f}$ to denote the coefficient of $r^{-k}(\ln r)^n$ in the large-$r$ expansion of a function $f$, as in
\begin{align}
f(u,r,\hat{x})=\sum_{k}\sum_{n}\frac{(\ln r)^{n}}{r^{k}}\,\overset{k,n}{f}(u,\hat{x}).
\end{align}
In the early and late time expansion, we will also encounter both inverse powers of $u$ and powers of logarithms of $u$.
In this context, we use a similar notation $f^{(m,n)}$, as in
\begin{align}
f(u,\hat{x})=\sum_{m}\sum_{n}\frac{(\ln u)^{n}}{u^{m}}\hspace{1mm}f^{(m,n)}({\hat{x}})
.
\end{align}
The free matter current $j^\text{free}_{u}=O(r^{-2})$ sources logarithmic terms in the gauge field, 
\begin{align}
A_{u}(u,r,\hat{x})&\overset{r\to\infty}{=}\frac{\ln r}{r}\overset{1,1}{A_{u}}(u,\hat{x})+\frac{1}{r}\overset{1,0}{A_{u}}(u,\hat{x})+\cdots,
\end{align}
through the Maxwell's equation
\begin{align}
    \overset{2,0}{j_{u}}=2\partial_{u}\overset{1,1}{A_{u}}.
\end{align}
Note that the $\overset{1,1}{A_u}$ component is overleading compared to the $1/r$ behavior of the free field. The Lorenz gauge condition \eqref{Lorenzscri} relates
\begin{align}
    \partial_{u}\overset{2,1}{A_{r}}=-\overset{1,1}{A_{u}}
    \label{Lorenz_ArAu}
    ,
\end{align}
from which it follows that 
\begin{align}
A_{r}(u,r,\hat{x})\overset{r\to\infty}{=}\frac{\ln r}{r^{2}}\overset{2,1}{A_{r}}(u,\hat{x})+\cdots
.
\end{align}
The gauge field corrected by the free matter current is
\begin{align}
A_{u}(u,r,\hat{x})&=\frac{\ln r}{r}\overset{1,1}{A_{u}}(u,\hat{x})+\frac{1}{r}\overset{1,0}{A_{u}}(u,\hat{x})+\cdots
\label{Round-1, leading A_u r-falloff}
,\\
A_{r}(u,r,\hat{x})&=\frac{\ln r}{r^{2}}\overset{2,1}{A_{r}}(u,\hat{x})+\cdots \label{Round-1, leading A_r r-falloff}
,\\
A_{C}(u,r,\hat{x})&=\overset{0,0}{A_{C}}(u,\hat{x})+\cdots \label{Round-1, leading A_C r-falloff}
.
\end{align}
Note that while the free fields are $\overset{k,0}{A}=O(e^0)$, the overleading Coulombic modes are $\overset{k,1}{A}=O(e)$. 

%%%%%%%%%%%%%%%%%%%%%%%%%%%%%%%%%%
\subsubsection*{Dressed matter}
%%%%%%%%%%%%%%%%%%%%%%%%%%%%%%%%%%

Through the scalar field equation \eqref{Field Equations for phi and A}, the Coulombic gauge modes feed back into the scalar field altering the large-$r$ falloff via dressing. We will use the notation
\begin{equation}\label{Asymptotic scalar large-r ansatz}
\phi(u,r,\hat{x})\overset{r\to\infty}{=}\sum_{k}\sum_{n}\frac{(\ln r)^{n}}{r^{1+k}}\phi_{k,n}(u,\hat{x}).
\end{equation}
The large-$r$ corrections to the free Klein-Gordon equation due to the radiative and the Coulombic gauge field have the following structure
\begin{align}
\Big[\nabla^{2}+\frac{1}{r}\Big(2ie\overset{1,0}{A_{u}}+2ie\overset{1,1}{A_{u}}\ln r+\cdots\Big)\partial_{r}+\frac{1}{r^{2}}\Big(2ie\overset{2,0}{A_{r}}+2ie\overset{2,1}{A_{r}}\ln r +\cdots\Big)\partial_{u}+\cdots\Big]\phi=0.
\end{align}
Because the correction to the Klein-Gordon equation is subleading compared to the free equation, the interacting scalar field is still dominated by the free behaviour, $\phi(x)=\phi^{\rm free}(x)+o(1/r)$, 
which implies $\phi_{0,n}=\delta_{n,0}\phi_{0,0}
$ in \eqref{Asymptotic scalar large-r ansatz}. This behavior is in stark contrast to the situation in massive QED \cite{Choi:2024mac} where Coulombic modes give rise to a logarithmically divergent dressing for the massive scalar field. In massless QED, the matter dressing by Coulombic gauge modes only comes in at subleading order 
\begin{align}\label{phisubdressed}
\phi(u,r,\hat{x})
\overset{r\to\infty}{=}\frac{1}{r}\phi^1_{\rm free}(u,\hat{x})+\sum_{n\geq 0}\frac{(\ln r)^{n}}{r^{2}}\phi_{1,n}(u,\hat{x})+\dots
\end{align}
with the explicit form for $\phi_{1,n}$ given in appendix \ref{Appendix on the dressed scalar field}.

The large-$r$ behaviour of this dressed matter field determines the asymptotic form of the interacting matter current
\begin{equation}
\label{large r falloffs of the dressed current}
\badat{3}
j_{u}(u,r,\hat{x})&\overset{r\to\infty}{=}\frac{1}{r^{2}}\overset{2,0}{j_{u}}(u,\hat{x})+\frac{\ln r}{r^3}\overset{3,1}{j_{u}}(u,\hat{x})+\frac{1}{r^{3}}\overset{3,0}{j_{u}}+\cdots,
\\
j_{r}(u,r,\hat{x})&\overset{r\to\infty}{=}\frac{\ln r}{r^{4}}\overset{4,1}{j_{r}}(u,\hat{x})+\frac{1}{r^{4}}\overset{4,0}{j_{r}}(u,\hat{x})+\cdots,
\\
j_{C}(u,r,\hat{x})&\overset{r\to\infty}{=}\frac{1}{r^{2}}\overset{2,0}{j_{C}}(u,\hat{x})+\frac{\ln r}{r^{3}}\overset{3,1}{j_{C}}(u,\hat{x})+\cdots\,.
\eadat
\end{equation}
These interacting matter currents in turn give rise to new Coulombic modes at higher orders in the coupling, which we return to in section \ref{Superphaserotation charge section}.

%%%%%%%%%%%%%%%%%%%%%%%%%%%%%%%%%%
\section{Superphaserotation symmetry}
\label{SuperphaserotationSymmetry}
%%%%%%%%%%%%%%%%%%%%%%%%%%%%%%%%%%
Gauge transformations
\begin{equation}
\delta A_{\mu}=\partial_{\mu}\epsilon, \quad \delta\phi=ie\epsilon\phi,
\end{equation}
are referred to as {\it large} if they give rise to non-vanishing Noether charges on the asymptotic phase space of the theory. To avoid any potential confusion with gauge redundancies, we follow \cite{Choi:2024mac} and refer to these physical transformations by how they act on matter fields, namely as a local enhancement of global phase rotations, or, {\it superphaserotations}.

In the Lorenz gauge the symmetry parameter $\epsilon(x)$ associated to superphaserotations satisfies the wave equation
\begin{align}
\nabla^{2}\epsilon=0.
\end{align}
The solution to this equation that is relevant for our discussion is the one that diverges linearly in $r$ \cite{Campiglia:2016hvg},
\begin{equation}\label{O(r) gauge parameter in Bondi coordinates}
\epsilon\left(u,r,\hat{x}\right)=r\varep+\frac{u}{2}\big(D^{2}+2\big)\varep+O\Big(\frac{\ln(r)}{r}\Big),
\end{equation}
where $\varep$ is a free function on $S^{2}$. 
Asymptotic charges with such divergent symmetry parameters have been found in \cite{Campiglia:2016hvg} to account for Low's subleading tree-level soft photon theorem \cite{Low:1958sn}. Accounting for the long-range interactions, it has been shown \cite{Choi:2024mac} that the same symmetry transformation governs the logarithmic soft photon theorem in massive scalar QED. It is thus natural to assume that the same symmetry that governs the massless limit of Low's soft theorem is also responsible for the logarithmic soft photon theorem in massless scalar QED.

%%%%%%%%%%%%%%%%%%%%%%%%%%%%%%%%%%
\section{Superphaserotation charge}\label{Superphaserotation charge section}
%%%%%%%%%%%%%%%%%%%%%%%%%%%%%%%%%%

The action of massless scalar QED is
\begin{equation}\label{Action of theory in form language}
S=\int\Big[-\frac{1}{2}F\wedge\star{F}-\Dcal\phi^{*}\wedge\star{\Dcal\phi}\Big],
\end{equation}
where $\Dcal$ refers to the gauge covariant derivative,
\begin{equation}
\Dcal\phi=\text{d}\phi-ieA\wedge\phi.
\end{equation}
The symplectic potential can be derived to be 
\begin{equation}
\Theta=-(\star{F})\wedge\delta A-\star{\Dcal\phi}\wedge\delta \phi^{*}-\star{\Dcal\phi^{*}}\wedge\delta \phi.
\end{equation}
It will be convenient to redefine our fields as
\begin{equation}\label{phiAredefine}
\phi\mapsto e^{ie\Lambda}\phi,\quad 
A\mapsto A+\text{d}\Lambda.
\end{equation}

Under a superphaserotation \eqref{O(r) gauge parameter in Bondi coordinates} parameterised by $\epsilon$ we find that $\Lambda$ changes as $\Lambda\mapsto \Lambda +\epsilon$, while the new matter and gauge fields are invariant and transform as usual under small gauge transformations. With the redefinition of fields \eqref{phiAredefine} the symplectic potential changes as 
\begin{align}
\Theta_{\text{old}}\mapsto\Theta= \Theta_{\text{old}}-\delta \Lambda\wedge\star{j}-\delta(\text{d}\Lambda)\wedge\star{F}.
\end{align}
The symplectic form $\Omega_{\Sigma}(\delta,\delta_\epsilon)=\int_{\Sigma} \left[\delta \Theta(\delta_\epsilon)-\delta_\epsilon \Theta(\delta)\right]$, with one of the variations, $\delta_\epsilon$, chosen to be a superphaserotation with parameter \eqref{O(r) gauge parameter in Bondi coordinates}, is
\begin{equation} \label{Symplectic form with symmetry parameter}
\Omega_{\Sigma}(\delta,\delta_{\epsilon})=-\int_{\Sigma}\epsilon\, \delta(\star{j})+\text{d}\epsilon\,\wedge\delta (\star{F}),
\end{equation}
where $\Sigma$ is a Cauchy surface. Since $\epsilon$ is field independent, it is a phase space scalar. This allows us to observe that the  expression \eqref{Symplectic form with symmetry parameter} can be written as a total variation.
\begin{equation}
\Omega_{\Sigma}(\delta,\delta_{\epsilon})=-\delta \int_{\Sigma}\left[\epsilon\left(\star j\right)+\text{d}\epsilon\wedge\left(\star F\right)\right]\eqqcolon-  \delta Q\left[\epsilon\right],
\end{equation}
We can extract the superphaserotation charge
\begin{equation}\label{QSPR}
Q\left[\epsilon\right]=\int_{\Sigma}\left[\epsilon\star j+\text{d}\epsilon\wedge\star F\right].
\end{equation}

To make the connection with soft theorems, it is useful to identify the  `soft' and `hard' contributions to this charge \eqref{QSPR}, i.e. the contributions that will map onto the soft photon insertion in the S-matrix and the contributions coming from the superphaserotation of the matter fields which will map onto the soft factor.
Interestingly, unlike in massive scalar QED, the term $\text{d}\epsilon\wedge\star{F}$ in \eqref{QSPR} turns out to contribute not just to the soft charge but also to the hard charge. This is already the case for the charge constructed in \cite{Campiglia:2016hvg}, whose conservation was shown to be equivalent to the subleading tree-level soft photon theorem in massless scalar QED. Here we will find a similar result for the logarithmic soft photon theorem.

While we focus on the future boundary a similar analysis can be carried out at the past boundary; we will use a subscript $\pm$ to distinguish future and past.
Taking $\Sigma$ to be a constant $t=u+r$ Cauchy slice, the expression for the charge in retarded Bondi coordinates is given by
\begin{equation}  
Q_+\left[\epsilon\right]=\int_{\Sigma}d^3x \hspace{1mm}r^{2}\hspace{1mm}\bigg[\epsilon\hspace{0.5mm}j_{u}+\partial_{r}\epsilon \hspace{0.5mm}F_{ur}-\partial_{u}\epsilon\hspace{0.5mm}F_{ur}+\frac{1}{r^{2}}D^{A}\epsilon \hspace{0.5mm}F_{uA}\bigg]
.
\label{Pre fall-off substitution charge expression}
\end{equation}
Substituting the large-$r$ fall-offs of the gauge field \eqref{Round-1, leading A_u r-falloff}-\eqref{Round-1, leading A_C r-falloff} and matter current \eqref{large r falloffs of the dressed current} we find
\begin{align}
Q_+\left[\vare\right]&=\int_{\Sigma}d^3x\hspace{1mm} \bigg[\underbrace{(u+r)\vare\overset{2,0}{j_{u}}}_{Q^{(-1)}_H}+\underbrace{\frac{u}{2}\big(D^{2}\vare\big)\overset{2,0}{j_{u}}+\vare\overset{3,0}{j_{u}}}_{ Q^{(0)}_H}+\underbrace{\ln r \,\vare\overset{3,1}{j_{u}}}_{ Q^{(\ln)}_H}\nonumber
\\  &\hspace{0.8cm}\underbrace{-(u+r)\vare D^{A}\overset{0,0}{F_{uA}}}_{ Q^{(-1)}_S}\hspace{2mm}\underbrace{-\vare\frac{1}{2}D^{2}\overset{2,0}{F_{ur}}-\vare D^{A}\overset{1,0}{F_{uA}}}_{ Q^{(0)}_H}\underbrace{-\vare\frac{u}{2}D^{2}D^{A}\overset{0,0}{F_{uA}}}_{ Q^{(0)}_S,\, Q^{(\ln)}_S}\bigg].
\label{FullCharge}
\end{align}
Here we highlighted the different types of charges that each term contributes to.

At large $|u|$, the leading term of the angular gauge field component, $\overset{0,0}{A_{C}}=A_{C}$, which enters in the field strengths in \eqref{FullCharge} has the following behaviour:
\begin{equation}\label{ACufalloff}
A_{C}(u,\hat{x})\overset{u\to\pm\infty}{=} A_{C}^{(0,0),\pm}(\hat{x})+\sum_{n=1}^{\infty}\frac{(\ln|u|)^{n}}{u}A_{C}^{(1,n),\pm}(\hat{x})+\frac{1}{u}A_{C}^{(1,0),\pm}(\hat{x})+O\left(\frac{1}{|u|^{1+\#}}\right).
\end{equation}
As in the massive QED analysis \cite{Choi:2024mac} it turns out the tower of logarithmic terms in \eqref{ACufalloff} vanishes. 
Thus, the charge \eqref{FullCharge}, computed from the symplectic structure is linearly divergent in $r$ and $u$, logarithmically divergent in $r$, and potentially logarithmically divergent in $|u|$, as arising from the $1/|u|$ term in \eqref{ACufalloff}. To regulate the logarithmic divergences at large $r$ and large-$|u|$ we introduce a cutoff\footnote{Note that choosing different cutoffs to regulate large-$r$ and large-$u$ divergences only affects the charge corresponding to the subleading tree-level soft theorem. This ambiguity is expected: logarithmic terms in the soft expansion render the power-law tree-level expansion (beyond the Weinberg pole) ambiguous.} at $\ln\Lambda^{-1}$, similar to the case of massive QED \cite{Choi:2024mac}. In the following we will show that the three charges, $Q^{(-1)}$, $Q^{(0)}$ and $Q^{(\ln)}$, that we can extract from \eqref{FullCharge} are associated to, respectively, the leading Weinberg, subleading Low, and logarithmic Sahoo-Sen soft theorems.

%%%%%%%%%%%%%%%%%%%%%%%%%%%%%%%%%%
\subsection{Tree-exact leading charge $Q^{(-1)}$}
%%%%%%%%%%%%%%%%%%%%%%%%%%%%%%%%%%

The leading superphaserotation charge $Q^{(-1)}_+$ corresponding to the term linearly divergent in time $t=u+r$ is
\begin{equation}
Q^{(-1)}_{+}\left[\vare\right]=\int_{\Sigma}\text{d}u\text{d}^{2}\Omega_{2}\hspace{2mm}\vare\Big[\overset{2,0}{j_{u}}-D^{A}\overset{0,0}{F_{uA}}\Big].
\end{equation}
We recognize this expression as the future contribution to the conserved charge that reproduces Weinberg's leading soft photon theorem.
The matter current contributes to the hard charge
\begin{equation}
Q^{(-1)}_{H,+}\left[\vare\right]=\int_{\Sigma}\text{d}u\text{d}^{2}\Omega_{2}\hspace{2mm}\vare \overset{2,0}{j_{u}},
\end{equation}
where
\begin{equation}
    \overset{2,0}{j_{u}}=ie(\phi_{0,0}\partial_{u}\phi_{0,0}^{*}-\phi_{0,0}^{*}\partial_{u}\phi_{0,0})
    .
\end{equation}

The soft charge is given by
\begin{align}
Q^{(-1)}_{S,+}[\vare]=-\int_{\Sigma}\text{d}u\text{d}^{2}\Omega_{2}\hspace{1.5mm}\vare D^{A}\overset{0,0}{F_{uA}}=\int_{S^{2}}\text{d}^{2}\Omega_{2}\hspace{1.5mm}\vare\mathcal{J}^{(-1)}_{+},
\end{align}
where $\mathcal{J}^{(-1)}_{+}$ is the leading soft photon current
\begin{equation}
\mathcal{J}^{(-1)}_{+}=-\int_{-\infty}^{\infty}\text{d}u\hspace{1.5mm}\partial_{u}(D^{C}\overset{0,0}{A_{C}})\hspace{5mm}\text{with}\hspace{5mm}\overset{0,0}{A_{C}}\overset{u\to+\infty}{=}A_{C}^{(0),+}+O\big(\frac{1}{u^\#}\big),
\end{equation}
with $\#$ any positive number and where 
\begin{equation}
\overset{0,0}{A_{C}}\overset{u\to+\infty}{=} \frac{1}{4\pi}\int\text{d}u'\text{d}^{2}\Omega_{2}' \hspace{2mm}\frac{D_{C}q\cdot q'}{q\cdot q'}\overset{2,0}{j_{u}}(u',\hat{x}').
\end{equation}

%%%%%%%%%%%%%%%%%%%%%%%%%%%%%%%%%%
\subsection{Loop-exact logarithmic charge $Q^{(\ln)}$}
%%%%%%%%%%%%%%%%%%%%%%%%%%%%%%%%%%
The logarithmic superphaserotation charge $Q^{(\ln)}_+$ is extracted from the logarithmically divergent terms in~\eqref{FullCharge}, namely from
\begin{align}\label{Qlnfinal}
\int_{\Sigma}\text{d}u\text{d}^{2}\Omega_{2}\hspace{1mm}\left[\ln r\, \vare\overset{3,1}{j_{u}}-\frac{1}{2}\left(D^{2}\vare\right) u\partial_{u}D^{C}\overset{0,0}{A_{C}}\right].
\end{align}
Regulating the large-$r$ divergence in \eqref{Qlnfinal} with the infrared cutoff $\Lambda^{-1}$, its coefficient gives the logarithmic hard charge
\begin{equation}\label{Logarithmic hard charge}
Q^{(\ln)}_{H,+}=\int_{\Sigma}\text{d}u\text{d}^{2}\Omega_{2}\hspace{1mm}\vare\overset{3,1}{j_{u}}.
\end{equation}
By direct computation, we show that in massless QED $\overset{3,1}{j_{u}}=0$. The details of this computation are given in appendices \ref{Appendix on the dressed scalar field} and \ref{Section on the current components formed out of the dressed scalar field}. Thus the hard logarithmic charge vanishes,
\begin{equation}
Q^{(\ln)}_{H,+}=0.
\end{equation}

To determine the soft charge, we express it in terms of the logarithmic soft photon current
\begin{equation}
\mathcal{J}^{(\ln)}_{+}=\int_{-\infty}^{+\infty}\text{d}u\hspace{1mm}\partial_{u}\big(u^{2}\partial_{u}D^{C}\overset{0,0}{A_{C}}\big)\hspace{5mm}\text{with}\hspace{5mm}\overset{0,0}{A_{C}}\overset{u\to+\infty}{=}A_{C}^{(0),+}+\frac{1}{u}A^{(1),+}_{C}+O\big(\frac{1}{u^{1+\#}}\big),
\end{equation}
with $\#$ any positive number, where the $1/u$ fall-off leads to a logarithmic divergence which we regulate with the same infrared cutoff $\Lambda^{-1}$. This gives the infrared-regulated logarithmic soft charge
\begin{equation}
Q^{(\ln)}_{S,+}=-\frac{1}{2}\int_{S^{2}}\text{d}^{2}\Omega_{2}\hspace{1mm}\left(D^{2}\vare\right)\mathcal{J}^{(\ln)}_{+}.
\end{equation}
To extract the coefficient of the $1/u$ term we start from the gauge field expressed in Cartesian coordinates,
\begin{equation}\label{Formal solution to Maxwell equations}
A_{\mu}(x)=\frac{1}{2\pi}\int\text{d}^{4}x'\sqrt{-g'}\hspace{1.5mm}\theta(x^{0}-x'^{0})\delta\big((x-x')^{2}\big)\hspace{1mm}j_{\mu}(x'),
\end{equation}
take the large-$r'$ limit and substitute the expressions for the currents.
This yields
\begin{align}
A_{C}^{(1,n),+} &=\frac{1}{4\pi}\int\text{d}u'\text{d}^{2}\Omega'_{2}\sum_{l=n}^{\infty}\binom{l}{n}\big[-\ln(-q\cdot q')\big]^{l-n}D_{C}q^{\mu}\big(-q'_{\mu}\overset{3,l}{j_{u}}+(D^{A}q'_{\mu})\overset{2,l}{j_{A}}\big)(u',\hat{x}').
\end{align}
Using $\overset{3,l}{j_{u}}=\overset{3,0}{j_{u}}\delta_{l,0}$ and $\overset{2,l}{j_{A}}=\overset{2,0}{j_{A}}\delta_{l,0}$ (see appendix \ref{Section on the current components formed out of the dressed scalar field}) together with current conservation, specifically  $\partial_{u}\overset{4,0}{j_{r}}-\overset{3,0}{j_{u}}-D^{C}\overset{2,0}{j_{C}}=0$, we immediately see that 
\begin{equation}\label{A1+is0}
A^{(1,0),+}_{C}\equiv A_{C}^{(1),+}=\frac{1}{4\pi}\Bigg[\int_{S^{2}}\text{d}^{2}\Omega_{2}'(-D_{C} q\cdot q')\overset{4,0}{j_{r}}\Bigg]^{u=+\infty}_{u=-\infty}=0,
\end{equation}
which vanishes because no current survives at the boundary of $\mathcal{I}^{+}$.
We also have
\begin{equation}
\forall n\geq 1:A_{C}^{(1,n),+}=0,
\end{equation}
which we anticipated above.
The logarithmic soft charge thus also vanishes
\begin{align}
Q^{(\ln)}_{S,+}=0.
\end{align}
This result turns out to be exact in the electromagnetic coupling as we show in appendix \ref{Appendix on exactness in coupling}.

%%%%%%%%%%%%%%%%%%%%%%%%%%%%%%%%%%
\subsection{Loop-corrected subleading charge $Q^{(0)}$}
%%%%%%%%%%%%%%%%%%%%%%%%%%%%%%%%%%
The subleading superphaserotation charge $Q^{(0)}_+$ corresponding to the terms in \eqref{FullCharge} that are finite at large distances is determined by
\begin{equation}\label{Subleading charge, having projected out Weinberg}
Q_+^{(0)}=\int_{\Sigma}\text{d}u\text{d}^{2}\Omega_{2}\hspace{1mm}\Big[\frac{u}{2}D^{2}\vare\,\overset{2,0}{j_{u}}+\vare\overset{3,0}{j_{u}}-\vare\left(\frac{1}{2}D^{2}\overset{2,0}{F_{ur}}+D^{A}\overset{1,0}{F_{uA}}+\frac{1}{2}u\partial_{u}D^{2}D^{C}\overset{0,0}{A_{C}}\right)\Big].
\end{equation}
Current conservation allows us to recast the second term as 
\begin{align}
\int_{\Sigma}\text{d}u\text{d}^{2}\Omega_{2}\hspace{1mm}\vare\overset{3,0}{j_{u}}=\int_{\Sigma}\text{d}u\text{d}^{2}\Omega_{2}\hspace{1mm}\vare \big(-D^{A}\overset{2,0}{j_{A}}\big)+\int_{\Sigma}\text{d}u\text{d}^{2}\Omega_{2}\hspace{1mm}\vare\overset{3,1}{j_{u}}+\Big[\int_{S^{2}}\text{d}^{2}\Omega_{2}\hspace{1mm}\vare\overset{4,0}{j_{r}}\Big]\Big|^{u=+\infty}_{u=-\infty}.
\end{align}
The second term on the rhs vanishes because the current component $\overset{3,1}{j_{u}}$ constructed from $\phi_{0,n}$ and $\phi_{1,n}$ is zero, while the third term vanishes because there are no currents at the boundaries of $\mathcal{I}^{+}$ in a massless theory. The remaining contributions to the charge can be expressed as 
\begin{equation}\label{Q0remaining}
\badat{2}
Q_+^{(0)}=&\int_{\Sigma}\text{d}u\text{d}^{2}\Omega_{2}\hspace{1mm}\left[\frac{u}{2} D^{2}\vare\, \overset{2,0}{j_{u}}-\varep D^{A}\overset{2,0}{j_{A}}-\vare\left(\frac{1}{2}D^{2}\overset{2,0}{F_{ur}}+D^{A}\overset{1,0}{F_{uA}}+\frac{1}{2}u\partial_{u}D^{2}D^{C}\overset{0,0}{A_{C}}\right)\right].
\eadat
\end{equation}

The Bianchi identity $\partial_{[u}F_{rA]}=0$ at $O\left(r^{-2}\right)$ along with the Maxwell equations for $\overset{3,0}{j_{u}}$, $\overset{3,1}{j_{u}}$ and $\overset{4,0}{j_{r}}$ allow us to write
\begin{equation}\label{Bianchi Identity [urA]}
\frac{1}{2}D^{2}\overset{2,0}{F_{ur}}+D^{A}\overset{1,0}{F_{uA}}=-\frac{1}{2}D^{A}\overset{2,0}{j_{A}}.
\end{equation}
Substituting \eqref{Bianchi Identity [urA]} back into \eqref{Q0remaining}, and rewriting the expression in terms of the vector field  \eqref{Vector field parameter for subleading soft theorem} gives
\begin{align}\label{Q0final}
Q_+^{(0)}=\int_{\Sigma}\text{d}u\text{d}^{2}\Omega_{2}\hspace{1mm}\left[u\big(D\cdot Y\big)\overset{2,0}{j_{u}}+Y^{A}\overset{2,0}{j_{A}}-\big(D\cdot Y\big)u\partial_{u}D^{C}\overset{0,0}{A_{C}}\right],
\end{align}
where, for comparison with the established literature, we defined the following vector field on $S^{2}$,
\begin{equation}\label{Vector field parameter for subleading soft theorem}
Y^{A}(\hat{x})\coloneqq\frac{1}{2}D^{A}\varep.
\end{equation}

Introducing the subleading soft photon current
\begin{align}\label{J0+}
\mathcal{J}^{(0)}_{+}&=\int_{-\infty}^{+\infty}\text{d}u\hspace{1mm}u\partial_{u}D^{C}\overset{0,0}{A_{C}},\quad \overset{0,0}{A_{C}}\overset{u\to +\infty}{=}A_{C}^{(0),+}+O\big(\frac{1}{u^{1+\#}}\big),
\end{align}
where $\#$ is any positive number,
we see that the last term in \eqref{Q0final} corresponds precisely to the soft charge associated to Low's subleading soft photon theorem  
\begin{align}
Q^{(0)}_{S,+}[Y]&=-\int_{\Sigma}\text{d}u\text{d}^{2}\Omega_{2}\,(D\cdot Y) u\partial_{u}D^{C}\overset{0,0}{A_{C}}=-\int_{S^{2}}\text{d}^{2}\Omega_{2}\,(D\cdot Y)\mathcal{J}^{(0)}_{+}.
\end{align}

Meanwhile, the first two terms in \eqref{Q0final} define the hard charge
\begin{equation}\label{LowHardCharge}
Q_{H,+}^{(0)}[Y]=\int_{\Sigma}\text{d}u\text{d}^{2}\Omega_{2}\hspace{1mm}\left[u\big(D\cdot Y\big)\overset{2,0}{j_{u}}+Y^{A}\overset{2,0}{j_{A}}\right].
\end{equation}
The covariant form of this expression matches the one obtained in \cite{Campiglia:2016hvg} whose conservation between past and future boundary were shown to be equivalent to Low's subleading soft photon theorem. While their expression for the current $\overset{2,0}{j_{A}}$ only includes free matter contributions, our expression accounts for infrared corrections from the coupling between the massless matter and the gauge field,
\begin{equation}
  \overset{2,0}{j_{A}}=ie(\phi_{0,0}\partial_{A}\phi^{*}_{0,0}-\phi^{*}_{0,0}\partial_{A}\phi_{0,0})-2e^{2}|\phi_{0,0}|^{2}\overset{0,0}A_{A} .
\end{equation}
This $O(e^2)$ contribution corresponds to a one-loop correction to Low's subleading soft photon theorem. For divergent superphaserotations with parameter \eqref{O(r) gauge parameter in Bondi coordinates} there are no further corrections at higher orders in the coupling.

%%%%%%%%%%%%%%%%%%%%%%%%%%%%%%%%%%
\section{Tail memory}
\label{TailMemory}
%%%%%%%%%%%%%%%%%%%%%%%%%%%%%%%%%%

The velocity kick memory is determined by the difference in the early and late time value of the gauge field
\begin{equation}
    \Delta A_C^{(0)}
    =
        A_C^{(0),+}-A_C^{(0),-},
\end{equation}
with $A_C^{(0),\pm}=A_{C}^{(0,0),\pm}(\hat{x})$ defined in \eqref{ACufalloff}. This `velocity kick' measures the shift in velocity of charged particles caused by the time-integrated electric field (the electric field tangential to the celestial sphere at infinity) \cite{Bieri:2013hqa}. It can either be a result of charged matter passing through $\mathcal{I}^{\pm}$, or through photons in an electromagnetic wave that passes through $\mathcal{I}^{\pm}$ redistributing the electric field over the celestial $S^{2}$. 
The long-range nature of the electromagnetic interactions lead to a {\it tail} in this memory given by
\begin{equation}
    \Delta A_C^{(1)}
    =
        A_C^{(1),+}-A_C^{(1),-}.
\end{equation}
From our result \eqref{A1+is0} it immediately follows that 
\begin{equation}
    \Delta A_C^{(1)}=0,
\end{equation}
which is exact in $e$.
The tail to the velocity kick memory thus vanishes for massless particles and to all orders in the electromagnetic coupling!

%%%%%%%%%%%%%%%%%%%%%%%%%%%%%%%%%%
\section{Conclusion}
\label{Conclusion}
%%%%%%%%%%%%%%%%%%%%%%%%%%%%%%%%%%

The long-range nature of electromagnetic interactions has been shown to modify the low-energy expansion of the S-matrix in scalar QED by logarithmic terms \cite{Sahoo:2018lxl} which can be translated to a late-time divergence in the radiation dressing the matter by a logarithmically divergent phase and the universality of these results is explained by the underlying superphaserotation symmetry \cite{Choi:2024mac}. The massless limit of the logarithmic soft photon theorem can immediately be seen to vanish (at least up to potential collinear divergences). This begs for an explanation in terms of symmetries. 

To arrive at the corresponding result of a vanishing superphaserotation charge, requires an entirely different analysis because the matter reaches not timelike but null infinity. This does not only lead to different asymptotic matter fall-offs but also turns on a Coulombic gauge field on null infinity, so that the gauge field has both radiative and Coulombic components on the same boundary.

In our analysis we assumed that the relevant symmetry underlying the logarithmic soft photon theorem in massless scalar QED is the same divergent superphaserotation that is responsible for Low's tree-level subleading soft photon theorem. This is motivated by the corresponding situation in scalar QED with massive matter. It is further supported by the fact that our derivation of the superphaserotation charge reproduces the charges associated to the leading and subleading tree-level soft photon theorems. The long-range interactions manifest themselves in logarithmic large-distance and late-time divergences in the hard and soft charge which have to be regulated. The coefficients of these infrared-regulated logarithmic terms define the logarithmic hard and soft charges which we show to vanish to all orders in the electromagnetic coupling. 

We furthermore compute the corrections to the charge associated to the subleading tree-level soft photon theorem in massless scalar QED to higher orders in the coupling. In the presence of logarithmic corrections this charge, and the corresponding soft theorem, would be ambiguous. The vanishing of the logarithmic charge, however, shows that it is not.

A further corollary of our result is that the tail to the velocity kick memory exactly vanishes to all orders in the coupling. This completes the classical infrared triangle for scalar QED with massless matter accounting for the long-range nature of the interactions: all three corners --- logarithmic soft photon theorem, charge conservation law associated to divergent superphaserotation symmetry and tail to the velocity kick memory --- all vanish exactly to all orders in the coupling!

While our asymptotic symmetry analysis focused on the symmetry interpretation of the classical logarithmic soft photon theorem which vanishes in massless scalar QED, there is a non-trivial quantum logarithmic soft photon theorem. It would be interesting to understand its symmetry interpretation and derive a quantum-corrected logarithmic charge. We leave this for future work.

\section*{Acknowledgements}

We would like to thank Alok Laddha and Prahar Mitra for insightful discussions. SC and AP are supported by the European Research Council (ERC) under the European Union’s Horizon 2020 research and innovation programme (grant agreement No 852386). This work was supported by the Simons Collaboration on Celestial Holography.

\appendix

\section{Dressed matter fields $\phi_{0,n}$ and $\phi_{1,n}$}\label{Appendix on the dressed scalar field}

We assume the following large-$r$ ansatz for the scalar field; a large-$r$ profile that has been altered from its `free version' due to long-range interactions:
\begin{align}
\phi(u,r,\hat{x})&=\sum_{k=0}^{\infty}\sum_{l=0}^{\infty}\frac{(\ln r)^{l}}{r^{1+k}} \phi_{k,l}(u,\hat{x}).
\end{align}
We have seen that the $k=0$ terms are trivial. This reduces the ansatz to
\begin{align}
\phi(u,r,\hat{x})&=\frac1r\phi_{0,0}(u,\hat{x})+\sum_{k=1}^{\infty}\sum_{l=0}^{\infty}\frac{(\ln r)^{l}}{r^{1+k}} \phi_{k,l}(u,\hat{x}).\label{Reduced Field Equations for phi and A}
\end{align}
The equation of motion for the scalar field itself \eqref{Reduced Field Equations for phi and A} can be expanded in a large $r$ series:
\begin{align}
\sum_{n=2}^{\infty}\sum_{m=0}^{\infty}\frac{(\ln r)^{m}}{r^{n}}\big(\text{E.O.M.}\big)_{n,m}=0.
\end{align}
The $n=2$ component of the E.O.M. is nothing but the free Klein-Gordon equation. The next goal is to solve this equation order by order in its large $r$ expansion. We begin with the equation of motion at $n=3$. This results in the following expression:
\begin{align}
0&=\sum_{n=0}^{\infty}\frac{(\ln r)^{n}}{r^{3}}\Big[-2(n+1)\partial_{u}\phi_{1,n+1}+2\partial_{u}\phi_{1,n}+2ie\big(\overset{2,1}{A_{r}}\partial_{u}-\overset{1,1}{A_{u}}\big)\phi_{0,0}\delta_{n,1}\nonumber
\\
&\hspace{2.5cm}+\big(D^{2}+2ie\overset{2,0}{A_{r}}\partial_{u}-2ie\overset{1,0}{A_{u}}-2ie\overset{0,0}{A^{C}}D_{C}-e^{2}\overset{0,0}{A^{C}}\overset{0,0}{A_{C}}\big)\phi_{0,0}\delta_{n,0}\Big].
\end{align}
The expression inside the bracket must vanish for every $n$. This leads to a recurrence relation for $\phi_{1,n}$. This recurrence relation then leads to the following solution, $\forall n\geq 0$:
\begin{align}
\partial_{u}\phi_{1,n}&=-\frac{1}{2n!}\Big[D^{2}-2ie\overset{1,0}{A_{u}}+2ie\overset{2,0}{A_{r}}\partial_{u}-e^{2}\overset{0,0}{A^{C}}\overset{0,0}{A_{C}}-2ie\overset{0,0}{A^{C}}D_{C}\Big]\phi_{0,0} \delta_{n,0}\nonumber
\\
&+\frac{ie}{n!}\Big[\overset{1,1}{A_{u}}-\overset{2,1}{A_{r}}\partial_{u}\Big]\phi_{0,0}(\delta_{n,0}+\delta_{n,1}).
\label{duphi1n}
\end{align}
This expression can be integrated along the null direction to get \hspace{1mm}\footnote{It is possible to get a $u$-independent (only $\hat{x}$-dependent) constant of integration. But we require $\phi\overset{u\to -\infty}{=}0.$ This sets that angle-dependent function to zero.}
\begin{align}
\phi_{1,0}&=-\frac{1}{2}\Big[\int_{-\infty}^{u}\text{d}u'\hspace{1mm}J(u',\hat{x})\phi_{0,0}(u',\hat{x})\Big]+\int_{-\infty}^{u}\text{d}u'\hspace{1mm}K(u',\hat{x})\phi_{0,0}(u',\hat{x})
,
\\
\phi_{1,1}&=\int_{-\infty}^{u}\text{d}u'\hspace{1mm}K(u',\hat{x})\phi_{0,0}(u',\hat{x})
,
\label{phi11_0}
\end{align}
where $J$ and $K$ are the following operators
\begin{align}
J(u,\hat{x})&=D^{2}+2ie\overset{2,0}{A_{r}}\partial_{u}-2ie\overset{1,0}{A_{u}}-e^{2}\overset{0,0}{A^{C}}\overset{0,0}{A_{C}}-2ie\overset{0,0}{A^{C}}D_{C}
,
\\
K(u,\hat{x})&=ie\big(\overset{1,1}{A_{u}}-\overset{2,1}{A_{r}}\partial_{u}\big).
\end{align}
Using the Maxwell equation for $\overset{2,0}{j_{u}}$ and the large $u$-falloffs of the free data $\phi_{0,0}$,  and the Lorenz gauge condition
\eqref{Lorenz_ArAu} which reads
\begin{equation}\label{Constraint on A_u(1,1) and A_r(2,1)}
\overset{1,1}{A_{u}}+\partial_{u}\overset{2,1}{A_{r}}=0
,
\end{equation}
the expression \eqref{phi11_0} for $\phi_{1,1}$ reduces to 
\begin{equation}
\phi_{1,1}(u,\hat{x})=\int_{-\infty}^{u}\text{d}u'\hspace{1mm}K(u',\hat{x})\phi_{0,0}(u',\hat{x})=-ie\overset{2,1}{A_{r}}(u,\hat{x})\phi_{0,0}(u,\hat{x}).
\label{phi_11}
\end{equation}

%%%%%%%%%%%%%%%%%%%%%%%%%%%%%%%%%%
\section{Perturbatively corrected matter current} \label{Section on the current components formed out of the dressed scalar field}
%%%%%%%%%%%%%%%%%%%%%%%%%%%%%%%%%%

In this appendix we derive, to cubic order in the coupling, the expressions for the matter current components that are used in the analysis of the main text.
These results also play a central role in the next section, where we derive an expression for the charge that is exact in the coupling.

The current is given by
\begin{align}
    j_{\mu}=ie(\phi\partial_{\mu}\phi^{*}-\phi^{*}\partial_{\mu}\phi)-2e^{2}A_{\mu}|\phi|^{2}
    .
    \label{jmu}
\end{align}
Expanding the right hand side using the expression \eqref{Reduced Field Equations for phi and A} for our scalar modes $\phi_{0,0}$ and $\phi_{1,n}$, and recalling from \eqref{duphi1n} that $\phi_{1,n}$ vanishes for all $n\geq 2$, we obtain the following results for the perturbatively corrected current components that are relevant to our analysis:
\begin{enumerate}
\item
    $\overset{2,n}{j_{u}}$ is zero for all $n\geq 1$, and $\overset{2,0}{j_u}$ is given by \begin{align}
        \overset{2,0}{j_{u}}=ie\phi_{0,0}\partial_{u}\phi^{*}_{0,0}+\text{c.c.}
    \end{align}
\item
    $\overset{3,n}{j_{u}}$ is zero for all $n\geq 1$.
    The case $n\geq 2$ follows immediately from \eqref{Reduced Field Equations for phi and A} and \eqref{jmu}.
    For $n=1$, the expression for $\overset{3,1}{j_u}$ is given by
    \begin{equation}
    \overset{3,1}{j_{u}}=\Big[ie\big(\phi_{0,0}\partial_{u}\phi^{*}_{1,1}+\phi_{1,1}\partial_{u}\phi^{*}_{0,0}\big)+\text{c.c.}\Big]-2e^{2}\overset{1,1}{A_{u}}|\phi_{0,0}|^{2}
    .
    \end{equation}
    Substituting the expression \eqref{phi_11} for $\phi_{1,1}$ we find that all terms cancel:
    \begin{align}\label{Vanishing of j_u(3,1)}
        \overset{3,1}{j_{u}}
        &=
            \Big[
            e^{2}\overset{1,1}{A_{u}}|\phi_{0,0}|^{2}-e^{2}\overset{2,1}{A_{r}}\phi_{0,0}\partial_{u}\phi^{*}_{0,0}+e^{2}\overset{2,1}{A_{r}}\phi_{0,0}\partial_{u}\phi^{*}_{0,0}
            +\text{c.c.}
            \Big]
            -2e^{2}\overset{1,1}{A_{u}}|\phi_{0,0}|^2
        = 0.
    \end{align}
    Since the hard charge $Q_H^{(\ln),+}$ given in \eqref{Logarithmic hard charge} is linear in $\overset{3,1}{j_u}$, it vanishes identically as a consequence of \eqref{Vanishing of j_u(3,1)}.

\item
    $\overset{3,n}{j_{r}}$ vanishes for all $n\geq0$, which follows immediately from \eqref{jmu}:
    \begin{align}
    \overset{3,n}{j_{r}}&=ie\phi_{0,0}\big(\partial_{r}\phi^{*}\big)_{1,n}+\text{c.c.}
    =
        -ie|\phi_{0,0}|^{2}\delta_{n,0}
        +\text{c.c.}
    = 0
    .
    \end{align}
    
\item
    $\overset{4,n}{j_{r}}$ is zero for all $n\geq 1$.
    The case $n\geq 2$ follows immediately from \eqref{Reduced Field Equations for phi and A} and \eqref{jmu}.
    For $n=1$, the current is given by
    \begin{align}
        \overset{4,1}{j_{r}}
        &=
            \Big[
            - ie(
                2\phi_{0,0}\phi^*_{1,1}
                + \phi_{1,1}\phi^*_{0,0}
            )
            +\text{c.c.}
            \Big]
            -2e^{2}\overset{2,1}{A_{r}}|\phi_{0,0}|^{2}.
    \end{align}

    Using the expression for $\phi_{1,1}$ given in \eqref{phi_11}, we find that all terms cancel out:
    \begin{align}
        \overset{4,1}{j_r}
        &=
            e^2\big(
            \overset{2,1}{A_{r}}|\phi_{0,0}|^{2}
            +\text{c.c.}
            \big)
            -2e^{2}\overset{2,1}{A_{r}}|\phi_{0,0}|^{2}
        = 0
        .
    \end{align}

\item
    $\overset{2,n}{j_{C}}$ vanishes for all $n\geq 1$,
    and $\overset{2,0}{j_{C}}$ is given by the expression
    \begin{align}
        \overset{2,0}{j_{C}}=ie\phi_{0,0}\partial_{C}\phi^{*}_{0,0}+\text{c.c.} -2e^{2}\overset{0,0}{A}_{C}|\phi_{0,0}|^{2}
        .
    \end{align}
\item
    $\overset{3,n}{j_{C}}$ vanishes for all $n\geq 2$.
    Again, the case $n\geq 2$ follows immediately from \eqref{Reduced Field Equations for phi and A} and \eqref{jmu}.
    For $n=1$, the expression for $\overset{3,1}{j_C}$ takes the form
    \begin{align}
        \overset{3,1}{j_C}
        &=
            \Big[
                ie
                \big(
                    \phi_{0,0}\partial_{C}\phi^*_{1,1}
                    +\phi_{1,1}\partial_{C}\phi^*_{0,0}
                \big)
                +\text{c.c.}
            \Big]
            -2e^{2}\overset{0,0}{A_{C}}
            \big(
                \phi_{0,0}\phi^*_{1,1}
                +\phi_{1,1}\phi^*_{0,0}
            \big)
            -2e^{2}\overset{1,1}{A_{C}}|\phi_{0,0}|^{2}
        .
    \end{align}
    Using the expression for $\phi_{1,1}$ from \eqref{phi_11}, and writing $\overset{3,1}{j_{C}}$ in terms of the field strength, we find again that all terms cancel out:
    \begin{align}
        \overset{3,1}{j_C}
        &=
        2e^{2}
        |\phi_{0,0}|^{2}  \overset{2,1}{F_{rC}}
        =0
        .
    \end{align}
    In the last equation we have used the fact that Maxwell equations set $\partial_{u}\overset{2,1}{F_{rC}}=0$, making $\overset{2,1}{F_{rC}}$ non-dynamical and allowing us to set it to zero.
\end{enumerate}

%%%%%%%%%%%%%%%%%%%%%%%%%%%%%%%%%%
\section{Exactness in coupling}\label{Appendix on exactness in coupling}
%%%%%%%%%%%%%%%%%%%%%%%%%%%%%%%%%%
Recall that the free matter current falls off at large $r$ as given by \eqref{j-free, r-falloff}. To the first order correction in the coupling, these leading free current components generate the following gauge field components
\begin{equation}
\label{Recap of leading Coulombic mode falloffs}
\badat{3}
A_{u}(u,\hat{x})&=\frac{\ln r}{r}\overset{1,1}{A_{u}}(u,\hat{x})+\frac{1}{r}\overset{1,0}{A_{u}}(u,\hat{x})+\cdots
\\
A_{r}(u,\hat{x})&=\frac{\ln r}{r^{2}}\overset{2,1}{A_{r}}(u,\hat{x})+\frac{1}{r^{2}}\overset{2,0}{A_{r}}(u,\hat{x})+\cdots
\\
A_{C}(u,\hat{x})&=\overset{0,0}{A_{C}}(u,\hat{x})+\cdots
\eadat
\end{equation}
These gauge field modes dress the scalar field. The dressed $\phi_{1,n}$ along with the free $\phi_{0,0}$\ \footnote{$\phi_{0,n}=\delta_{n,0}\phi_{0,0}$ because unlike the massive case, here we have no non-zero $\overset{1,0}{A_{r}}$.} gives us the perturbatively corrected matter-current. The leading large-$r$ falloffs of this current are (see appendix \ref{Section on the current components formed out of the dressed scalar field})
\begin{equation}
\label{Recap of corrected current r-falloffs}
\badat{3}
j_{u}(u,\hat{x})&=\frac{1}{r^{2}}\overset{2,0}{j_{u}}(u,\hat{x})+\frac{1}{r^{3}}\overset{3,0}{j_{u}}(u,\hat{x})+\cdots
,\\
j_{r}(u,\hat{x})&=\frac{1}{r^{4}}\overset{4,0}{j_{r}}(u,\hat{x})+\cdots
,\\
j_{C}(u,\hat{x})&=\frac{1}{r^{2}}\overset{2,0}{j_{C}}(u,\hat{x})+\cdots
.
\eadat
\end{equation}
Notice that these leading falloffs are the same as the ones for $j_{\mu}^{\text{free}}$. Because the $O(e)$ Coulombic modes \eqref{Recap of leading Coulombic mode falloffs} were determined purely by the leading free-current components, the  $O(e^{3})$ Coulombic modes that will be generated by the current components in \eqref{Recap of corrected current r-falloffs} will have the same leading large-$r$ behaviors as in \eqref{Recap of leading Coulombic mode falloffs}.
It follows that the expressions obtained in appendix \ref{Section on the current components formed out of the dressed scalar field} for $\phi_{0,n}$ and $\phi_{1,n}$, written in terms of the free data $\phi_{0,0}$ and the gauge field components, remain unchanged at higher orders in the coupling.
As a consequence, the conclusions drawn in the previous appendix continue to hold beyond the leading order correction.
Therefore, the statement that the charges $\overset{(\ln)}{Q_{H}}$ and $\overset{(\ln)}{Q_{S}}$ vanish remains true to all orders in the coupling.

Now we look at the charges associated with the order-$\omega^0$ soft theorem. Recall from equation \eqref{Pre fall-off substitution charge expression} that the exact expression for the charge is given by
\begin{equation} \label{Conserved Noether charge on a constant time slice}
Q\left[\epsilon\right]=\int_{\Sigma}\text{d}u\text{d}^{2}\Omega_{2}\hspace{1mm}\Big[r^{2}\epsilon j_{u}+r^{2}\partial_{r}\epsilon F_{ur}-r^{2}\partial_{u}\epsilon F_{ur}+\gamma^{AB}\partial_{A}\epsilon F_{uB}\Big]
\end{equation}
Given the large-$r$ falloffs of all the quantities in \eqref{Conserved Noether charge on a constant time slice}, the only surviving terms in the large-$r$ limit of \eqref{Conserved Noether charge on a constant time slice} involve the following quantities:
\begin{equation}
\big(\hspace{1mm}\overset{3,n}{j_{u}},\overset{2,n}{j_{u}},\overset{2,n}{F_{ur}},\overset{1,n}{F_{uA}},\overset{0,n}{F_{uA}}\hspace{1mm}\big)\nonumber
\end{equation}
At large $r$, the non-vanishing charges are
\begin{align}
Q[\epsilon]&=\int_{\Sigma}\text{d}u\text{d}^{2}\Omega_{2}\sum_{n=0}^{\infty}(\ln r)^{n}\Big[\vare r\overset{2,n}{j_{u}}+\vare \overset{3,n}{j_{u}}+\frac{u}{2}D^{2}\vare \overset{2,n}{j_{u}}+u\vare \overset{2,n}{j_{u}}+ \vare \overset{2,n}{F_{ur}}-\frac{1}{2}D^{2}\vare \overset{2,n}{F_{ur}}
\nonumber\\&\quad
-\vare\overset{2,n}{F_{ur}}-\vare r D^{A}\overset{0,n}{F_{uA}}-\vare D^{A}\overset{1,n}{F_{uA}}-\frac{u}{2}\vare D^{2}D^{A}\overset{0,n}{F_{uA}}-u\vare D^{A}\overset{0,n}{F_{uA}}\Big]. \label{Expresison for possible higher-logs in the conserved charges}
\end{align}
We have already argued that $\overset{2,n}{j_{u}}=\delta_{n,0}\overset{2,0}{j_{u}}$ and $\overset{3,n}{j_{u}}=\delta_{n,0}\overset{3,0}{j_{u}}$ to all orders in the coupling. We now need to look at $\{\overset{2,n}{F_{ur}},\overset{1,n}{F_{uA}},\overset{0,n}{F_{uA}}\}$. 
Given the large-$r$ fall off of the gauge field components at $O(e^{3})$, we have the following expressions,
\begin{equation}
\label{Field strength components in terms of the gauge field}
\badat{3}
\overset{2,n}{F_{ur}}&=\delta_{n,0}\big(\overset{1,0}{A_{u}}-\overset{1,1}{A_{u}}+\partial_{u}\overset{2,0}{A_{r}}\big)
,\\
\overset{0,n}{F_{uC}}&=\delta_{n,0}\big(\partial_{u}\overset{0,0}{A_{C}}\big)
,\\
\overset{1,n}{F_{uC}}&=\delta_{n,0}\big(\partial_{u}\overset{1,0}{A_{C}}-\partial_{C}\overset{1,0}{A_{u}}\big).
\eadat
\end{equation}
All the gauge field components in \eqref{Field strength components in terms of the gauge field} are generated by the leading current components\footnote{Apart from the leading $\overset{2,0}{j_{u}},\overset{4,0}{j_{r}},\overset{2,0}{j_{C}}$, the subleading $\overset{3,0}{j_{u}},\overset{3,0}{j_{C}}$ also play a role in generating some of these gauge field modes, but $\overset{3,0}{j_{u}},\overset{3,0}{j_{C}}$ have been shown to fall off exactly as the free current (i.e. $\overset{3,n}{j_{u}}=\overset{3,0}{j_{u}}\delta_{n,0}$ and $\overset{3,n}{j_{C}}=\overset{3,0}{j_{C}}\delta_{n,0}$, no overleading logs).}.
We have already shown that large-$r$ behaviour of these components does not change at higher order in the coupling.
Therefore,  \eqref{Expresison for possible higher-logs in the conserved charges} reduces to
\begin{equation}\label{Charge expredssion, now perturbatively exact}
Q[\epsilon]=\int_{\Sigma}\text{d}u\text{d}^{2}\Omega_{2}\hspace{1mm}\Big[t\vare\big(\overset{2,0}{j_{u}}-D^{A}\overset{0,0}{F_{uA}}\big)+\vare\overset{3,0}{j_{u}}+\frac{u}{2}D^{2}\vare\overset{2,0}{j_{u}}-\vare\big(\frac{1}{2}D^{2}\overset{2,0}{F_{ur}}+D^{A}\overset{1,0}{F_{uA}}+\frac{u}{2}D^{2}D^{A}\overset{0,0}{F_{uA}}\big)\Big].
\end{equation}
This result, which is now exact to all orders in the coupling, agrees with what we obtained in \eqref{FullCharge} (with the observation that $\overset{3,1}{j_{u}}=0$ now put in).

We therefore conclude that the expressions obtained for the order-$\omega^0$ charge and the (vanishing) logarithmic charge are perturbatively exact.

%%%%%%%%%%%%%%%%%%%%%%%%%%%%%%%%%%
\section{Absence of collinear divergences in the classical limit}\label{Appendix on collinear divergences}
%%%%%%%%%%%%%%%%%%%%%%%%%%%%%%%%%%
We are dealing with a theory where the massless gauge bosons interact with massless matter. Usually, in scattering processes, when analysing loops, such theories lead to collinear divergences. In going through the Feynman diagrammatic computation for the classical logarithmic soft photon theorem, one finds that the classical soft factor is
\begin{align}
S_{\text{classical,em}}^{\left(0\right)}&=\sum_{a,b\atop b\neq a}q_{a}^{2}q_{b}\frac{\varepsilon_{\mu}k_{\nu}}{(p_{a}\cdot k)}\left\{p_{a}^{\mu}\frac{\partial}{\partial p_{a\nu}}-p_{a}^{\nu}\frac{\partial}{\partial p_{a\mu}}\right\}\Bigg[i\hspace{1mm}\big(p_{a}\cdot p_{b}\big)\hspace{1mm}\mathcal{I}_{ab}\Bigg]. \label{Expression for S^(0),cl,em}
\end{align}
Here, $\{\varepsilon_{\mu},k_{\nu}\}$ are the polarization and four-momentum respectively of the external soft photon. $\{p_{a}^{\mu},q_{a}\}$ are the four-momentum and electric charge respectively of the $\text{a}^{\text{th}}$ hard particle. $\mathcal{I}_{ab}$ is the following loop integral:
\begin{align}
\mathcal{I}_{ab}&=\frac{i}{E_{a}E_{b}}\int_{\text{reg}}\frac{d^{3}\ell}{(2\pi)^{3}}\hspace{1.5mm}\frac{1}{\Big[\big(\frac{\vec{p}_{b}\cdot \vec{\ell}}{E_{b}}\big)^{2}-|\vec{\ell}|^{2}\Big]}\frac{1}{\Big[\big(\frac{\vec{p}_{b}}{E_{b}}-\frac{\vec{p}_{a}}{E_{a}}\big)\cdot \vec{\ell}-i\epsilon\Big]}.
\end{align}
The subscript `reg' on the integral indicates that the integral is to be taken over the region where $|\ell^{\mu}|$ is large compared to $\omega$ but small compared to the energies of the finite energy
particles, see \cite{Sahoo:2018lxl} for details on the explicit evaluation of the integrals. Parameterizing the $\ell$-space in spherical coordinates, with $\vec{p}_{b}$ along the Z-axis and the XY-projection of $\vec{p}_{a}$ along the X-axis, calling the `fixed' angle between $\vec{p}_{a}$ and $\vec{p}_{b}$ $\alpha_{ab}$, this integral becomes
\begin{align}
\mathcal{I}_{ab}&=\frac{i}{E_{a}E_{b}}\frac{\ln\omega^{-1}}{8\pi^{3}}\int_{-1}^{1}{\rm d}\cos \theta_{\ell}\int_{0}^{2\pi}\text{d}\varphi_{\ell}\frac{1}{(\cos ^{2}\theta_{\ell}-1)\big(-\sin \theta_{\ell}\cos \varphi_{\ell}\sin \alpha_{ab}+\cos \theta_{\ell}(1-\cos \alpha_{ab})-i\epsilon\big)}.
\end{align}
Define 
\begin{align}
A(\theta_{\ell},\alpha_{ab})&\coloneqq\cos \theta_{\ell}(1-\cos \alpha_{ab}).
\\
B(\theta_{\ell},\alpha_{ab})&\coloneqq 
\sin \theta_{\ell}\sin \alpha_{ab}.
\end{align}
With this,
\begin{align}
\mathcal{I}_{ab}&=\frac{i}{E_{a}E_{b}}\frac{\ln\omega^{-1}}{8\pi^{3}}\int_{-1}^{1}\text{d}\cos \theta_{\ell}\frac{1}{\big(\cos ^{2}\theta_{\ell}-1\big)}\int_{0}^{2\pi}\text{d}\varphi_{\ell}\hspace{2mm}\frac{1}{A(\theta_{\ell},\alpha_{ab})-B(\theta_{\ell},\alpha_{ab})\cos (\varphi_{\ell})-i\epsilon}.
\end{align}
The $\theta_{\ell}$ integral can be divided up into five regions, each one potentially contributing some divergence. These five regions are  when $\cos \theta_{\ell}\sim \pm 1$, when $-\cos (\frac{\alpha_{ab}}{2})<\cos \theta_{\ell}<\cos (\frac{\alpha_{ab}}{2})$ and when $\cos \theta_{\ell}\sim\pm\cos (\frac{\alpha_{ab}}{2})$. If we cutoff the $\theta_{\ell}$ integral at $\pm 1$ with the same regulator, say $\delta$ then it is easy to see that the contributions coming from these two regulated regions cancel each other exactly. $\delta$ can then safely be taken to zero. In the region where $-\cos (\frac{\alpha_{ab}}{2})<\cos \theta_{\ell}<\cos (\frac{\alpha_{ab}}{2})$, the $\varphi_{\ell} $ integrand develops simple poles and a $\pm i\pi \delta(x)+P.V.(1/x)$ prescription can be used to perform the $\varphi_{\ell}$ integral. The $\theta_{\ell}$ integral in this region then gives the following contribution
\begin{equation}
\mathcal{I}_{ab}^{|\cos \theta_{\ell}|<\cos (\frac{\alpha_{ab}}{2})}=-\frac{1}{4\pi}\ln(\omega^{-1})\frac{1}{\left(p_{a}\cdot p_{b}\right)}.
\end{equation}
In the regions where $\cos \theta_{\ell}\sim\pm\cos (\frac{\alpha_{ab}}{2})$, say regulated by $\tilde{\delta}$ such that $|\cos \theta_{\ell}-\cos (\frac{\alpha_{ab}}{2})| <\tilde{\delta}$ for $\cos \theta_{\ell}\sim +\cos (\frac{\alpha_{ab}}{2})$ and  $|\cos \theta_{\ell}+\cos (\frac{\alpha_{ab}}{2})| <\tilde{\delta}$ for $\cos \theta_{\ell}\sim -\cos (\frac{\alpha_{ab}}{2})$, the integrand diverges either at $\varphi_{\ell}=0$ or at $\varphi_{\ell}=\pi$ respectively. In this case, even though the integrand seems to diverge, the divergence is soft enough for the final $\theta_{\ell}$ integral to scale as $\tilde{\delta}^{1/2}$. In the limit where we take the cutoff $\tilde{\delta}$ to zero, the contributions from these two regions vanish. So finally, the integral $\mathcal{I}_{ab}$ is only contributed to by the region where $-\cos (\frac{\alpha_{ab}}{2})<\cos \theta_{\ell}<\cos (\frac{\alpha_{ab}}{2})$:
\begin{equation}
\mathcal{I}_{ab}=-\frac{1}{4\pi}\ln(\omega^{-1})\frac{1}{\left(p_{a}\cdot p_{b}\right)}.
\end{equation}
The expression for the classical logarithmic soft factor, \eqref{Expression for S^(0),cl,em}, immediately tells us that 
\begin{equation}
S_{\text{classical,em}}^{\left(0\right)}=-\frac{i}{4\pi}\ln(\omega^{-1})\sum_{a,b\atop b\neq a}q_{a}^{2}q_{b}\frac{\varepsilon_{\mu}k_{\nu}}{(p_{a}\cdot k)}\left\{p_{a}^{\mu}\frac{\partial}{\partial p_{a\nu}}-p_{a}^{\nu}\frac{\partial}{\partial p_{a\mu}}\right\}\Big[1 \Big]=0.
\end{equation}
This shows that as far as the classical logarithmic soft factor in massless sQED is concerned, it vanishes without suffering from any collinear divergences.
%%%%%%%%%%%%%%%%%%%%%%%%%%%%%%%%%%%%%%%%%%%%%%%%%%%%%%%%%%%%

\printbibliography
\end{document}